\documentclass[transmag]{IEEEtran}
\usepackage[utf8]{inputenc}
\usepackage{latexsym}
\usepackage{graphicx}
\usepackage{amsfonts,amssymb,amsmath}
\usepackage{hyperref}
\usepackage[neverdecrease]{paralist}
\usepackage{amsmath}
\usepackage{color}
\usepackage{algorithm}
\usepackage{tikz}
\usepackage{graphicx}
\usepackage{authblk}
\usepackage{mathtools}
\usepackage{amsfonts}
\usepackage{amssymb}
\usepackage{enumitem}
\usepackage{amsthm}

\usepackage[noadjust]{cite}
\usepackage{dblfloatfix}
\usepackage{array,multirow}

\theoremstyle{definition}
\newtheorem{definition}{Definition}

\newtheorem{remark}{\bf Remark}

\newtheorem{theorem}{Theorem}

\newtheorem{proposition}{Proposition}

\makeatletter
\renewcommand*\env@matrix[1][\arraystretch]{%
	\edef\arraystretch{#1}%
	\hskip -\arraycolsep
	\let\@ifnextchar\new@ifnextchar
	\array{*\c@MaxMatrixCols c}}
\makeatother

\DeclareGraphicsExtensions{.eps,.png}

\def\BibTeX{{\rm B\kern-.05em{\sc i\kern-.025em b}\kern-.08em T\kern-.1667em\lower.7ex\hbox{E}\kern-.125emX}}
\begin{document}

\title{Distributed Mixed Voltage Angle and Frequency Droop Control of Microgrid Interconnections with \\Loss of Distribution-PMU Measurements}

\author{S. Sivaranjani\textsuperscript{*}, Etika Agarwal\textsuperscript{*}, Vijay Gupta, Panos Antsaklis, and Le Xie{\textsuperscript{\textdagger}}
\thanks{\textsuperscript{*}These authors contributed equally to this work. {\textsuperscript{\textdagger}} Corresponding author.}
\thanks{S. Sivaranjani and Le Xie were supported in part by NSF grants ECCS-1839616, ECCS-1611301 and CCF-1934904, and the Department of Energy Contract DE-EE0009031. S. Sivaranjani was also partially supported in this work by the Schlumberger Foundation Faculty for the Future fellowship. Vijay Gupta was partially supported by NSF grant CNS-1739295 and DARPA FA8750-20-2-0502. Etika Agarwal and Panos Antsaklis were partially supported by the Army Research Office Grant No. ARL W911NF-17-1-0072.}
\thanks{S. Sivaranjani and Le Xie are with the Department of Electrical and Computer Engineering, Texas A\&M University, College Station, TX; \{sivaranjani,le.xie\}@tamu.edu. Etika Agarwal is with GE Global Research, India; \{etika.agarwal09@gmail.com\}. S. Sivaranjani and Etika Agarwal were with the Department of Electrical Engineering, University of Notre Dame, Notre Dame, IN when this work was carried out. Vijay Gupta and Panos Antsaklis are with the Department of Electrical Engineering, University of Notre Dame, Notre Dame, IN; \{vgupta2,pantsakl\}@nd.edu.}}

\IEEEtitleabstractindextext{\begin{abstract}Recent advances in distribution-level phasor measurement unit (D-PMU) technology have enabled the use of voltage phase angle measurements for direct load sharing control in distribution-level microgrid interconnections with high penetration of renewable distributed energy resources (DERs). In particular, D-PMU enabled voltage angle droop control has the potential to enhance stability and transient performance in such microgrid interconnections. However, these angle droop control designs are vulnerable to D-PMU angle measurement losses that frequently occur due to the unavailability of a global positioning system (GPS) signal for synchronization. In the event of such measurement losses, angle droop controlled microgrid interconnections may suffer from poor performance and potentially lose stability.  In this paper, we propose a novel distributed mixed voltage angle and frequency droop control (D-MAFD) framework to improve the reliability of angle droop controlled microgrid interconnections. In this framework, when the D-PMU phase angle measurement is lost at a microgrid, conventional frequency droop control is temporarily used for primary control in place of angle droop control to guarantee stability. We model the microgrid interconnection with this primary control architecture as a nonlinear switched system and design distributed secondary controllers to guarantee  stability of the network. Further, we incorporate performance specifications such as robustness to generation-load mismatch and network topology changes in the distributed control design. We demonstrate the performance of this control framework by simulation on a test 123-feeder distribution network. \end{abstract}


}

\maketitle

\section{INTRODUCTION}
\label{sec:intro}
	\IEEEPARstart{P}{hasor} measurement units (PMUs) have been extensively used in real-time wide-area monitoring, protection and control (WAMPAC) applications at the transmission level. However, in traditional distribution networks with one-way power flows and no active loads, real-time monitoring and control using phasor measurements has not been necessary for reliable operation \cite{sexauer2013phasor}. Further, small angle deviations and consequently, poor signal-to-noise ratios, make real-time estimation of voltage phase angles at the distribution level an extremely challenging problem. Therefore, applications of distribution-level PMUs (D-PMUs) have so far been confined to the substation-level, with angle references being synchronized with the transmission network \cite{von2014micro}. However, in future distribution systems, new architectures like microgrids with large-scale integration of renewable distributed energy resources (DERs) and flexible loads, as well as new economic paradigms like demand response, will require extensive real-time monitoring and control. { In this context, D-PMUs (such as the $\mu$PMU) that can provide accurate measurements of small angle deviations ($\approx0.001^\circ$) and voltage magnitude deviations ($\approx$ 0.0002 p.u.) with high sampling rates ($\approx120s^{-1}$) have recently been developed \cite{von2017precision}-\nocite{peisert2017lbnl}\cite{jamei2017anomaly}, and are expected to be critical components of future power distribution infrastructure \cite{sexauer2013phasor}\cite{sanchez2013current}.}
	
		In particular, consider the problem of ensuring stability and reliability in distribution networks comprised of interconnected microgrids. Typically, such microgrid interconnections are controlled in a hierarchical manner with three layers of control \cite{guerrero2011hierarchical}\cite{vasquez2010hierarchical} - (i) a primary control layer to ensure proper load sharing between microgrids, (ii) a secondary control layer to ensure system stability by eliminating frequency and voltage deviations, and (iii) a tertiary control layer to provide power reference set points for individual microgrids. The primary control layer commonly comprises of frequency droop and voltage droop controllers to regulate the real and reactive power outputs respectively of each microgrid at the point of common coupling (PCC). These droop control characteristics are implemented artificially using voltage-source inverter (VSI) interfaces that are designed such that individual microgrids emulate the dynamics of synchronous generators. 
	
	However, the use of frequency droop controllers in networks with a large penetration of low inertia VSI-interfaced microgrids has been demonstrated to result in numerous issues including chattering \cite{majumder2010improvement}, loss of synchronization, and undesirable frequency deviations resulting from the trade-off between active power sharing and frequency accuracy \cite{guerrero2009control}. With the availability of highly accurate D-PMUs, angle droop controllers that directly use the voltage phase angle deviation measurements at the PCC for active power sharing have been proposed to replace frequency droop controllers in the primary layer. These angle droop designs have been demonstrated to result in increased stability, smaller frequency deviations and faster dynamic response \cite{majumder2009angle,majumder2010power,zhang2015online,zhang2016transient,zhang2016interactive}. 
	
	A key bottleneck in the widespread adoption of D-PMU based control designs for microgrids is the reliability of D-PMU voltage phase angle measurements. Phase angle measurements from D-PMUs require a global positioning system (GPS) signal for synchronization of the angle reference across the network \cite{sexauer2013phasor}. However, recent studies have demonstrated that PMU GPS signals are frequently lost due to factors like weather events and communication failure, leading to loss of PMU measurements \cite{yao2016impact}\cite{yao2017gps}. In fact, such PMU measurement losses have been observed to occur as often as 6-10 times a day, with each loss event ranging from an average of 6-8 seconds to over 25 seconds \cite{huang2016data}. In WAMPAC applications, loss of GPS signal for PMUs has been demonstrated to result in severe performance degradation \cite{almas2016impact}. Furthermore, control strategies that rely on PMU measurements have been demonstrated to be vulnerable to GPS spoofing attacks, where a falsified GPS signal may be fed to compromise the PMU angle reference, leading to potentially catastrophic outcomes like cascade failures \cite{shepard2012evaluation}\cite{nercreport}. Therefore, in the event of D-PMU measurement losses, microgrid interconnections that rely solely on angle droop control will be particularly prone to poor dynamic performance and potential instability \cite{sivaranjani2018mixed}. 
	
	In wide-area control applications, the issue of PMU measurement loss is typically handled from a networked control systems perspective, where a popular approach is to allow  the controller to use the last available measurement in the event of a measurement loss, and bound the maximum allowable duration of the loss to guarantee stability \cite{sivaranjani2013networkedisgt}\cite{sivaranjani2013networked}. Typically, networked control designs also assume knowledge of the probability distributions of the loss events \cite{singh2015stability}. However, these approaches suffer from two critical issues in the context of distribution systems. { First, in D-PMUs, the duration of measurement loss may exceed the maximum allowable duration to guarantee stability.  This is due to the fact that angle droop controllers are extremely fast acting \cite{majumder2009angle}, and may therefore not be able to maintain stability \cite{sivaranjani2018mixed} under the longer measurement loss durations (typically tens of seconds) seen in D-PMU GPS signal losses \cite{huang2016data}. Second, even for measurement loss durations smaller than this threshold, large voltage angle and frequency drifts can occur due to the controller repeatedly using the incorrect (last available) measurement \cite{sivaranjani2018mixed}.} 
	
	In order to address stability and performance issues resulting from D-PMU measurement losses in angle droop controlled microgrid interconnections, we introduce the idea of mixed voltage angle and frequency droop control (MAFD). When D-PMU voltage angle measurements are lost at a microgrid due to loss of a GPS signal, frequency measurements may still be available, since they can be obtained locally without a GPS signal for synchronization. In the MAFD framework, frequency droop control is, therefore, temporarily used in place of angle droop control for primary control of active power sharing at particular microgrids where D-PMU measurements are lost \cite{sivaranjani2018mixed}. In a network of microgrids under the MAFD primary control framework, each microgrid may operate with either angle droop  or frequency droop control at any time instant, depending on the availability of D-PMU measurements at that microgrid. We therefore model the time-varying dynamics of a network of microgrids with MAFD primary control as a nonlinear switched system. We then show that the MAFD framework, along with a dissipativity-based distributed secondary controller, guarantees stability of angle droop controlled microgrid interconnections without any restrictions on the duration or probability distribution of D-PMU measurement losses. We refer to this control architecture with MAFD primary control and the proposed distributed secondary control as the distributed MAFD (D-MAFD) framework. Additionally, the D-MAFD design incorporates performance specifications including robustness to disturbances and network topology changes, with the view of increasing the role of D-PMU measurements in islanding operations and plug-and-play architectures in future microgrid interconnections. Finally, we show through case studies that the  D-MAFD framework significantly enhances system stability in D-PMU measurement loss scenarios under conditions of generation-load mismatches and is robust to network topology changes induced by faults. { The key contributions of this paper are:
		\begin{itemize}[leftmargin=0.3cm]
			\item First, we develop a distributed control framework to guarantee stability of a network of microgrids operating under angle droop control, when D-PMU angle measurements are lost. We accomplish this via an MAFD primary control framework, where each microgrid can switch  from the default angle droop control mode to a frequency droop control mode if D-PMU angle measurements are lost at that microgrid. In contrast to the networked control systems approaches are typically adopted in literature to handle measurement losses \cite{sivaranjani2013networkedisgt}\cite{sivaranjani2013networked}, the D-MAFD framework does not impose any bounds on the duration or probability distribution of measurement loss to guarantee stability. Rather, it exploits the physics of the system (that is, the fact that the frequency is the derivative of the angle) to indirectly control the voltage angle using the frequency when angle measurements are lost. Therefore, the voltage angle can be controlled continuously without any loss of information even under D-PMU measurement losses.
			\item Second, we design a dissipativity-based distributed secondary controller to guarantee stability of the network of microgrids, when each microgrid can arbitrarily switch between angle and frequency droop control based on the availability of D-PMU angle measurements. This design extends the passivity-based state-feedback control design for nonlinear discrete-time switched systems in \cite{agarwal2017feedback} to a  general continuous-time output-feedback dissipativity framework, incorporating additional performance specifications and robustness to network topology changes (such as line outages).
	\end{itemize}}

	A preliminary version of this work was presented in \cite{sivaranjani2018mixed}. In this paper, we significantly expand the results in \cite{sivaranjani2018mixed} by incorporating robustness and performance specifications, additional case studies, and detailed proofs of all mathematical results that were omitted in \cite{sivaranjani2018mixed}.
	
		\begin{figure*}[b]
		\setcounter{figure}{1}
		\centering
		\includegraphics[scale=0.8,trim=2.5cm 0.4cm 19cm 0.5cm]{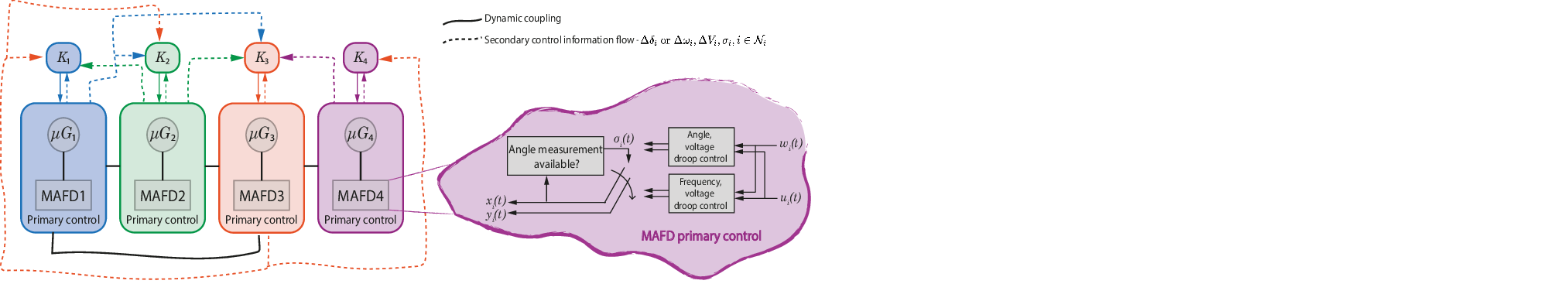}
		\caption{{Schematic of representative microgrid in the D-MAFD framework with MAFD primary control and distributed secondary control.}}
		\label{fig:dmafd_schematic}
	\end{figure*}
	\vspace{0.2em}
	\textit{Notation:} Let $\mathbb{R}$ and $\mathbb{R}^{n}$ denote the sets of real numbers and $n$-dimensional real vectors respectively. The $(i,j)$-th element of a matrix $A \in \mathbb{R}^{m\times n}$ is denoted by $A_{ij}$ and the transpose is denoted by $A'\in \mathbb{R}^{n \times m}$. The identity matrix is represented by $I$, with dimensions clear from the context. A symmetric positive (negative) definite matrix $P \in \mathbb{R}^{n \times n}$ is represented by $P>0$ ($P<0$).

	\section{Mixed Voltage Angle and Frequency Droop Control (MAFD) Model}
\label{sec:mafd}
Consider a network of $N$ microgrids where each microgrid is connected to the network at the PCC { as shown in Fig. \ref{fig:microgrids_schematic}.} Define $\mathcal{N}_i$ to be the \textit{neighbor set} of the $i$-th microgrid, that is, the set of all microgrids to which the $i$-th microgrid is directly connected in the network. For convenience of notation, we also include $i$ in this set. Considering the coupled AC power flow model, the real and reactive power injections $P_{inj}^j(t)$ and $Q_{inj}^j(t)$, at the $j$-th microgrid at time $t$ are given by
{\small\begin{equation}\label{powerflow}
	\begin{aligned}
	P_{inj}^j(t)&= \sum \limits_{k \in \mathcal{N}_j} V_j(t) V_k(t) |Y_{jk}| \sin(\delta_{jk}(t)+\pi /2-\angle{Y_{jk}}) \\
	Q_{inj}^j(t)&= \sum \limits_{k \in \mathcal{N}_j} V_j(t) V_k(t) |Y_{jk}| \sin(\delta_{jk}(t)-\angle{Y_{jk}}),
	\end{aligned}
	\end{equation}}where $V_j(t)$ and $\delta_j(t)$ are the voltage magnitude and phase angle at PCC respectively, $\delta_{jk}(t)=\delta_j(t)-\delta_k(t)$, and $Y_{jk}$ is the complex admittance of the line between the PCC buses of microgrids $j$ and $k$. 

{ As shown in Fig. \ref{fig:microgrids_schematic},} the primary control layer for every microgrid comprises of an angle droop and a voltage droop control loop, which regulate the real and reactive power injections of the microgrid respectively to track desired reference values $V_i^{ref}$ and $\delta_i^{ref}$ of the voltage magnitude and phase angle at the PCC respectively. {	These angle droop controllers can be readily implemented by programming the VSI-interfaces at the microgrid PCC \cite{majumder2010improvement}\cite{majumder2010droop}. Alternatively, these angle droop controllers may be implemented via angle feedback control in inertia emulating interfaces like virtual synchronous generators \cite{zhang2016clean}.  }
As mentioned in Section \ref{sec:intro}, the use of angle droop primary control to directly regulate real power has several advantages such as increased frequency stability and power sharing accuracy \cite{majumder2009angle}. The implementation of primary angle droop control schemes requires real-time angle measurements from D-PMUs placed at the microgrid PCCs, which in turn require a GPS signal for synchronization. However, since GPS signals are frequently lost, microgrid interconnections that primarily use angle droop control will suffer from poor performance and potential instability. 
\begin{figure}[t]
	\setcounter{figure}{0}
	\centering
	\includegraphics[scale=0.48,trim=0cm 0cm 0cm 0cm]{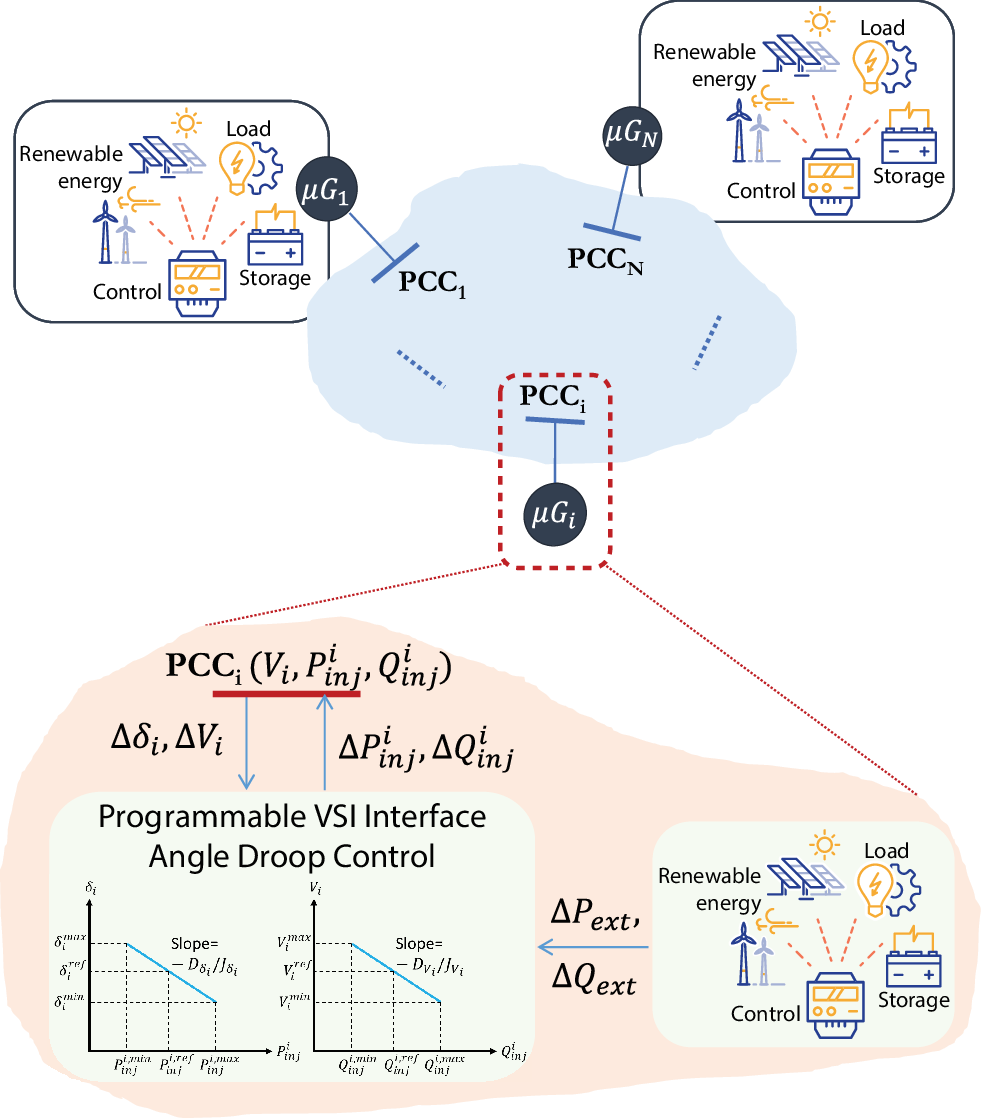}
	\vspace{-5pt}
	\caption{{Network of interconnected microgrids with angle droop control. Some icons are licensed from Adobe Stock and modified for this illustration.}}
	\label{fig:microgrids_schematic}
	\vspace{-1em}
\end{figure}	

To address this issue, when D-PMU angle measurements are lost at certain microgrids due to loss of a GPS signal loss, we employ a mixed angle and frequency droop control (MAFD) framework for primary control, as shown in Fig. \ref{fig:dmafd_schematic}. When D-PMU angle measurements are lost, frequency measurements can still be obtained locally at each microgrid, without the need for a GPS signal-based synchronization. In the MAFD framework, classical frequency droop control is therefore used in place of the angle droop control to temporarily regulate real power at those microgrids until D-PMU measurements are restored. Thus, at any given time, some microgrids may operate with angle droop control while others operate with frequency droop control depending on the availability of D-PMU measurements. At every time $t $, each microgrid in the MAFD framework operates in one of two modes, denoted by $\sigma_i(t)\in\{1,2\}$ - (i) angle droop control mode $(\sigma_i(t) =1)$, when real-time angle measurements are available from the D-PMU at that microgrid, and (ii) frequency droop control mode $(\sigma_i(t) = 2)$, when D-PMU voltage angle measurements are lost or corrupted at that microgrid due to GPS signal loss or sensor malfunction (Fig. \ref{fig:dmafd_schematic}). The dynamics of the $i\text{-th}$ microgrid in each of these modes is described as follows.

\vspace{2mm}
\noindent\textbf{Angle Droop Control Mode, $\mathbf{\sigma_i(t)=1}$:}
In this mode, D-PMU angle measurements are available, and the microgrid operates with angle and voltage droop control laws given by
\vspace{-1em}
{	\begin{align}
	J_{\delta_i}\Delta \dot \delta_i(t) &= -D_{\delta_i} \Delta \delta_i(t) + \Delta P_{ext}^i(t) -\Delta P_{inj}^i(t)\label{angle1}\\
	J_{V_i} \Delta \dot V_i(t) &= -D_{V_i} \Delta V_i(t) +\Delta Q_{ext}^i(t) -\Delta Q_{inj}^i(t), \label{voltage-angledroop}
	\end{align}}where $\Delta V_i(t)=V_i(t)-V_i^{ref}$ and $\Delta \delta_i(t)=\delta_i(t)-\delta_i^{ref}$, are the deviations of the PCC voltage magnitude from their reference values, $\Delta P_{inj}^i(t)=P_{inj}^i(t)-P_{inj}^{i,ref}$ and $\Delta Q_{inj}^i(t)=Q_{inj}^i(t)-Q_{inj}^{i,ref}$ are the deviations of the real and reactive power injections $P_{inj}^i(t)$ and $Q_{inj}^i(t)$ from their nominal values $P_{inj}^{i,ref}$ and $Q_{inj}^{i,ref}$ respectively, and $\Delta P_{ext}^i(t)$ and $\Delta Q_{ext}^i(t)$ are the generation-load mismatches at the $i$-th microgrid. The droop coefficients $J_{\delta_i}$, $D_{\delta_i}$, $J_{V_i}$ and $D_{V_i}$ can be implemented by programming the VSI interface at the PCC \cite{zhang2015online}. Additionally, the dynamics of the frequency error in the angle droop control mode is propagated as
{\begin{equation}\label{continuity}
	\begin{split}
	\Delta \dot \omega_i(t)=&-\frac{D_{\delta_i}}{J_{\delta_i}}\left[-\frac{D_{\delta_i}}{J_{\delta_i}}\Delta \delta_i(t)+\frac{1}{J_{\delta_i}}\Delta P_{ext}^i(t)\right. \\ &\left. -\frac{1}{J_{\delta_i}}\Delta P_{inj}^i(t)\right]-\frac{1}{J_{\delta_i}}\Delta \dot P_{inj}^i(t),
	\end{split}
	\end{equation}}where $\Delta \omega_i(t)=\omega_i(t)-\omega_i^{ref}$ is the deviation of the frequency of the $i$-th microgrid $\omega_i(t)$ from its reference value $\omega_i^{ref}$. The derivative $\Delta P_{inj}^i(t)$ is computed from \eqref{powerflow}.
{
	\begin{remark}\label{rem:continuity}
		We note that the propagation of the dynamics of the frequency error through \eqref{continuity} in the angle droop control mode is not carried out in typical angle droop control schemes. However, in our framework, this serves the important purpose of ensuring the continuity of the state $x_i(t)$ at every instant of time, thereby allowing for  seamless switching between the angle droop control and frequency droop control modes as follows. The continuity of the state can be guaranteed by ensuring that  \eqref{angle} is satisfied at every switching instant. While this is automatically ensured for switches from the frequency droop mode to the angle droop mode, \eqref{continuity} enforces $\Delta \dot \omega_i(t)=\Delta \ddot{\delta_i}(t)$ for all time~$t$, thereby ensuring continuity for switches from the angle droop mode to the frequency droop mode. Thus, through \eqref{continuity}, the continuity of the state $x_i(t)$ is guaranteed at every time instant for arbitrary switching between the angle and frequency droop control modes at the $i$-th microgrid.
\end{remark}	}

\vspace{2mm}
\noindent\textbf{Frequency Droop Control Mode, $\mathbf{\sigma_i(t)=2}$:} When D-PMU angle measurements are not available, a frequency droop control law is used to regulate the real power, and the dynamics of the microgrid in this mode are given by
{\begin{align}
	\Delta \dot \delta_i(t) &= \Delta \omega_i(t) \label{angle}\\
	J_{\omega_i}\Delta \dot \omega_i(t) &= -D_{\omega_i} \Delta \omega_i(t) + \Delta P_{ext}^i(t) -\Delta P_{inj}^i(t) \label{frequency}\\
	J_{V_i} \Delta \dot V_i(t) &= -D_{V_i} \Delta V_i(t) +\Delta Q_{ext}^i(t) -\Delta Q_{inj}^i(t), \label{voltage}
	\end{align}}
\noindent	where $J_{\omega_i}$ and $D_{\omega_i}$ are the frequency droop coefficients. 

We now define the state, input and disturbance vectors of the $i$-th microgrid to be $x_i(t)=[\Delta \delta_i(t) \; \Delta\omega_i(t) \; \Delta V_i(t)]'$, $u_i(t)=[\Delta P_{inj}^i(t) \quad \Delta Q_{inj}^i(t)]'$ and $w_i(t)=[\Delta P_{ext}^i(t) \quad \Delta Q_{ext}^i(t)]'$ respectively. The output vector of the $i$-th microgrid is 
\begin{equation}
y_i(t)=g_{\sigma_i(t)}^i(x_i(t),w_i(t)),
\end{equation}
where $g_{\sigma_i(t)}^i=[\Delta \dot \delta_i(t) \; \Delta V_i(t)]'$ when $\sigma_i(t)=1$ and $g_{\sigma_i(t)}^i(t)=[\Delta \dot \omega_i(t) \; \Delta V_i(t)]'$ when $\sigma_i(t)=2$. The dynamics of the $i$-th microgrid in the MAFD primary control framework can then be written as a nonlinear switched system

{\begin{equation}\label{switchedsystem_state}
	\begin{aligned}
	\dot{x}_i(t) & = f_{\sigma_i(t)}^i(x_i(t),u_i(t),w_i(t)) \\
	u_i(t) & = h^i(x_i(t)),
	\end{aligned}
	\end{equation}}where the dynamics $f^i_1(x_i(t),u_i(t),w_i(t))$ in the angle droop control mode are defined by \eqref{angle1}-\eqref{continuity}, the dynamics $f^i_2(x_i(t),u_i(t),w_i(t))$ in the frequency droop control mode are defined by \eqref{angle}-\eqref{voltage}, and $h^i(x_i(t))$ is given by the power flow model \eqref{powerflow} independent of the switching mode $\sigma_i$. Define the augmented state vector for the microgrid interconnection as $x(t){=}[x_1'(t), x_2'(t), \ldots, x_N'(t)]'$. Similarly, define the augmented input, disturbance and output vectors obtained by stacking the inputs, disturbances and outputs of all microgrids to be $u(t)$, $w(t)$ and $y(t)$ respectively. Finally, define the augmented switching vector $\sigma(t)=[\sigma_1(t),\cdots,\sigma_N(t)]'$, where every element can take values of 1 or 2, indicating the availability or loss of D-PMU angle measurements at that microgrid. Let $\Sigma$ denote the set of all possible values of this switching vector. We now write the dynamics of the microgrid interconnection with MAFD as the nonlinear switched system
\begin{subequations}\label{nonlinear_switched}
	\begin{equation}\label{nonlinear_switched_a}
		\begin{aligned}
		\dot{x}(t) & = f_{\sigma(t)}(x(t),u(t),w(t))\\
		y(t) & = g_{\sigma(t)}(x(t),w(t)) \\
		u(t) & = h(x(t)),
		\end{aligned}
		\end{equation}
		\vspace{-0.7em}
	{\small	\begin{align}\label{nonlinear_switched_b}
		f_{\sigma(t)}=\begin{bmatrix}
		f^1_{\sigma_1(t)} \\
		\vdots \\
		f^{N}_{\sigma_{N}(t)}
		\end{bmatrix}, \, g_{\sigma(t)}=\begin{bmatrix}
		g^1_{\sigma_1(t)} \\
		\vdots \\
		g^{N}_{\sigma_{N}(t)}
		\end{bmatrix}, \, h=\begin{bmatrix}
		h^1 \\
		\vdots \\
		h^{N}
		\end{bmatrix}.
		\end{align}}
\end{subequations}In order to design a secondary controller that guarantees the stability of the microgrid interconnection with MAFD, we linearize the system model \eqref{nonlinear_switched} around the origin to obtain a linearized switched system model
\begin{subequations}\label{linear switched system}
		\begin{equation}
		\begin{aligned}
		\dot{{x}}(t) & = A_{\sigma(t)}{x}(t) + B_{\sigma(t)}^{(1)}{u}(t) + B_{\sigma(t)}^{(2)}{w}(t)\\
		{y}(t)&  = C_{\sigma(t)}{x}(t) + D_{\sigma(t)}{w}(t)\\
		u(t) & =Hx(t),
		\end{aligned}
		\end{equation}
		\vspace{-0.7em}
	{\small		\begin{align}\label{linearization matrix}
		A_j & = \left.\frac{\partial f_j}{\partial x}\right\vert_{\substack{x=0\\w=0}}, & B_j^{(1)} & = \left.\frac{\partial f_j}{\partial u}\right\vert_{\substack{x=0\\w=0}},  & B_j^{(2)} & = \left.\frac{\partial f_j}{\partial w}\right\vert_{\substack{x=0\\w=0}}  \nonumber \\
		C_j & = \left.\frac{\partial g_j}{\partial x}\right\vert_{\substack{x=0\\w=0}}, & D_j & = \left.\frac{\partial g_j}{\partial w}\right\vert_{\substack{x=0\\w=0}}, & &
		\end{align}
		\begin{equation}
		{H = \begin{bmatrix}
			\frac{\partial u_1}{\partial x_1} & \cdots & \frac{\partial u_1}{\partial x_{N}}\\
			\vdots & \vdots &\vdots\\
			\frac{\partial u_{N}}{\partial x_1} & \cdots & \frac{\partial u_{N}}{\partial x_{N}}
			\end{bmatrix}}_{x=0},\end{equation}
		\begin{equation*}
		{\frac{\partial u_i}{\partial x_k} = \begin{bmatrix}
			\frac{\partial \Delta P^i_{inj}}{\partial \Delta \delta_k} & \frac{\partial \Delta P^i_{inj}}{\partial \Delta \omega_k} & \frac{\partial \Delta P^i_{inj}}{\partial \Delta V_k} \\[5pt] 
			\frac{\partial \Delta Q^i_{inj}}{\partial \Delta \delta_k} & \frac{\partial \Delta Q^i_{inj}}{\partial \Delta \omega_k} & \frac{\partial \Delta Q^i_{inj}}{\partial \Delta V_k}
			\end{bmatrix}}, \, {i,k \in \{1, \cdots , N\}}.\end{equation*}}
\end{subequations}Note that $H$ is the power flow Jacobian pertaining to the linearization of \eqref{powerflow}.
	\begin{figure*}[b]
	\normalsize
	\newcounter{MYtempeqncnt}
	\setcounter{MYtempeqncnt}{\value{equation}}
	\setcounter{equation}{13}
	\hrulefill	{\small\begin{subequations}\label{control_lmi}
			\begin{equation}
			\label{pass_lmi1}
			M_j = 
			\begin{bmatrix}[1.5]
			-P(A_j+B_j^{(1)}H)-(A_j+B_j^{(1)}H)'P-B_j^{(1)}U_jC_j-C_j'U_j'B_j^{(1)'} & -PB_j^{(2)}-B_j^{(1)}U_jD_j+C_j'S_j & -C_j'Q_{j-}^{1/2} \\
			-B_j^{(2)^{'}}P-D_j'U_j'B_j^{(1)^{'}}+S_j'C_j & D_j'S+S_j'D_j+R_j & -D_j'Q_{j-}^{1/2} \\
			-Q_{j-}^{1/2}C_j & -Q_{j-}^{1/2}D_j & I 
			\end{bmatrix}>0
			\end{equation}
			\begin{equation}\label{pass_lmi2}
			PB_j^{(1)} = B_j^{(1)}V_j, \quad Q_{j-}^{1/2}Q_{j-}^{1/2}=-Q_j
			\end{equation}
			\begin{equation}\label{pass_lmi3}
			V_j \in \mathcal{S}_v, \quad U_j \in \mathcal{S}_H	\vspace{-1em}
			\end{equation}
	\end{subequations}}
	\setcounter{equation}{\value{MYtempeqncnt}}
\end{figure*}

The MAFD primary control architecture allows for indirect control of the real power sharing in the microgrid interconnection in a decentralized manner even when D-PMU angle measurements are lost. With this architecture, { a supplementary controller, termed as the secondary controller, must then be designed to eliminate the voltage and angle errors that arise due to open-loop droop control.} In the following section, we propose a distributed secondary control design that uses only local information at each microgrid to regulate angle and voltage deviations and guarantee stability of the network of interconnected microgrids.

\section{D-MAFD Secondary Control Design} \label{sec:dmafd}
Consider the microgrid interconnection with MAFD primary control as shown in Fig. \ref{fig:dmafd_schematic}. Given the dynamics in \eqref{nonlinear_switched} and its linear approximation \eqref{linear switched system}, we would like to design a secondary controller that eliminates voltage and angle deviations in the microgrid interconnection and guarantees stability during D-PMU measurement losses. The proofs of all the results in this section are provided in the Appendix.
\subsection{Distributed Secondary Control Synthesis}	
In this section, we develop a distributed secondary control design for microgrid interconnections operating in the MAFD framework, where every microgrid locally determines its secondary control actions using information only from its immediate neighbors. The microgrid interconnection with MAFD primary control and distributed secondary control as shown in Fig. \ref{fig:dmafd_schematic} is termed as the \textit{distributed MAFD (D-MAFD)} framework. For this network, we would like to design a secondary output-feedback control law $\tilde u(t)=K_{\sigma(t)}y(t)$, ${K_j \in \mathbb{R}^{2N\times2N}}$, $j \in \Sigma$, such that the microgrid interconnection \eqref{nonlinear_switched} with $u(t) \mapsto u(t)+\tilde{u}(t)$ is $\mathcal{L}_2$-stable with respect to disturbances $w(t)$ even when D-PMU angle measurements are unavailable in any number of microgrids in the network, that is, \eqref{nonlinear_switched} switches arbitrarily between angle droop and frequency droop primary control modes of individual microgrids. Intuitively, the $\mathcal{L}_2$ stability property guarantees that the system outputs (angles, frequencies and voltages) are bounded for finite disturbances. 

{	In order to design a distributed secondary controller that guarantees the stability of the network of interconnected microgrids with MAFD with dynamics by \eqref{nonlinear_switched}, we will enforce the property of $QSR$-dissipativity, defined as follows. 
	\begin{definition}\label{def:QSR}
		The nonlinear switched system \eqref{nonlinear_switched} is said to be \textit{$QSR$-dissipative} with input $w$ and dissipativity matrices $Q_j$, $S_j$ and $R_j$, $j\in\Sigma$, if there exists a positive definite storage function $V(x):\mathbb{R}^{3N}\rightarrow\mathbb{R}_+$ such that,
		{{
				\begin{equation}
				\begin{bmatrix}
				y(t) \\
				w(t)
				\end{bmatrix}'\begin{bmatrix}
				Q_j & S_j\\
				S_j' & R_j
				\end{bmatrix}\begin{bmatrix}
				y(t) \\
				w(t)
				\end{bmatrix} \geq \dot{V}(x(t))\end{equation}
		}}holds for all $t \geq t_0 \geq 0$, where $x(t)$ is the state at time $t$ resulting from the initial condition $x(t_0)$ and input $w(\cdot)$. 
				\noindent Further, \eqref{nonlinear_switched} is said to be \textit{$QSR$-state strictly input dissipative} ($QSR$-SSID) if, for all $t\in \mathbb{R}_+$ and $j\in\Sigma$,  
		{\small{
				\begin{align}\label{QSR-SSID}
				\begin{bmatrix}
				y(t) \\
				w(t)
				\end{bmatrix}'{\begin{bmatrix}
					Q_j & S_j\\
					S_j' & R_j
					\end{bmatrix}}\begin{bmatrix}
				y(t) \\
				w(t)
				\end{bmatrix} {\geq} \dot{V}(x(t)) + \phi_j(w(t)) + \psi_j(x(t)),
				\end{align}}}where $\phi_j(\cdot),\psi_j(\cdot)$ are positive definite functions of $w(t)$ and $x(t)$ respectively. 
		Finally, a switched system \eqref{nonlinear_switched} is said to be \textit{locally $QSR$-dissipative} if it is $QSR$-dissipative for all $x\in\mathcal{X}$ and $w\in\mathcal{W}$ where $\mathcal{X}\times\mathcal{W}$ is a neighborhood of $(x,w)=0$.
	\end{definition}
	$QSR$-dissipativity is a very useful property for nonlinear switched systems, since it implies $\mathcal{L}_2$ stability as follows. 
	\begin{remark}\label{remark: QSR-L2}
		A $QSR$-dissipative switched system \eqref{nonlinear_switched} is $\mathcal{L}_2$ stable if $Q_j<0$ for every $j\in\Sigma$. 
	\end{remark}
	In addition to $\mathcal{L}_2$ stability, $QSR$-dissipativity can also be used to capture other useful dynamical properties such as robustness and transient performance via appropriate choice of the $Q_j$, $S_j$ and $R_j$ matrices \cite{agarwal2019compositional}. Such dissipativity-based controllers have also been shown to guarantee stability and performance in several microgrid applications \cite{agarwal2017feedback}\cite{sivaranjaniconic}. We now have the following result relating the dissipativity of the nonlinear system \eqref{nonlinear_switched} to that of its linear approximation \eqref{linear switched system}.

	\begin{proposition}\label{prop:linear_to_nonlinear}
		The nonlinear switched system \eqref{nonlinear_switched} is locally $QSR$-dissipative if its linear approximation \eqref{linear switched system} is $QSR$-SSID with the same dissipativity matrices and a quadratic storage function {{$V(x(t))=x(t)'Px(t)$}}, where {{$P\in\mathbb{R}^{3N \times 3N}$}} and {{$P>0$}}.
	\end{proposition}

	Proposition \ref{prop:linear_to_nonlinear} extends the results of \cite{agarwal2017feedback} to continuous-time switched systems. }Using the concept of $QSR$-dissipativity, we now show that a distributed secondary controller that guarantees $\mathcal{L}_2$ stability of the microgrid interconnection with MAFD primary control subject to disturbances $w(t)$ can be designed by solving linear matrix inequalities as follows.
\begin{figure*}[b]
	\normalsize
	\setcounter{equation}{16}
	\hrulefill
	{\footnotesize\begin{subequations}\label{pert_lmi}
			\begin{equation}
			\label{pert_lmi1}
			\begin{gathered}
			\hat{M}_j =
			\begin{bmatrix}[1.5]
			-P(A_j+B_j^{(1)}H)-(A_j+B_j^{(1)}H)'P-B_j^{(1)}U_jC_j-C_j'U_j'B_j^{(1)'} -2\gamma P & -PB_j^{(2)}-B_j^{(1)}U_jD_j+C_j'S_j & -C_j'Q_{j-}^{1/2} \\
			-B_j^{(2)^{'}}P-D_j'U_j'B_j^{(1)^{'}}+S_j'C_j & D_j'S+S_j'D_j+R_j & -D_j'Q_{j-}^{1/2} \\
			-Q_{j-}^{1/2}C_j & -Q_{j-}^{1/2}D_j & I 
			\end{bmatrix}>0 \\
			\end{gathered}
			\end{equation}
			\vspace{0.2em}
			\begin{equation}\label{pert_lmi2}
			PB_j^{(1)} = B_j^{(1)}V_j, \quad Q_{j-}^{1/2}Q_{j-}^{1/2}=-Q_j
			\end{equation}
			\begin{equation}\label{pert_lmi3}
			V_j \in \mathcal{S}_v, \quad U_j \in \mathcal{S}_H
			\end{equation}
	\end{subequations}}
\end{figure*}
\begin{theorem}\label{thm:design}
	If there exists symmetric positive definite matrix \(P>0\), negative definite matrix $Q_j<0$, and matrices \(U_j, V_j, S_j\) and $R_j$ of appropriate dimensions such that \eqref{control_lmi} is satisfied for every switching vector $j \in \Sigma$, then the distributed secondary control law $u(t) \mapsto u(t) + \tilde{u}(t)$ where \(\tilde{u}(t)=K_{\sigma(t)}y(t)\) with 
	\setcounter{equation}{14}
	\vspace{-0.5em}
	{\begin{equation} 
		\label{gain}
		K_j=V_j^{-1}U_j, \quad \forall j\in \Sigma,\vspace{-0.5em}
		\end{equation}}obtained by solving design equations \eqref{control_lmi}, where $\mathcal{S}_v$ is the set of all diagonal matrices and $\mathcal{S}_H$ is the set of all matrices with the same sparsity structure as the Jacobian matrix $H$ in \eqref{linear switched system}, is sufficient to guarantee $\mathcal{L}_2$ stability of the microgrid interconnection \eqref{nonlinear_switched} in the D-MAFD framework with respect to disturbances $w(t)$ under arbitrary loss of D-PMU angle measurements. 
\end{theorem}
\begin{figure*}[t]
	\setcounter{figure}{2}
	\centering
	\begin{minipage}{0.43\textwidth}
		\centering
		\includegraphics[scale=0.4,trim=0.5cm 0.5cm 0cm 0cm]{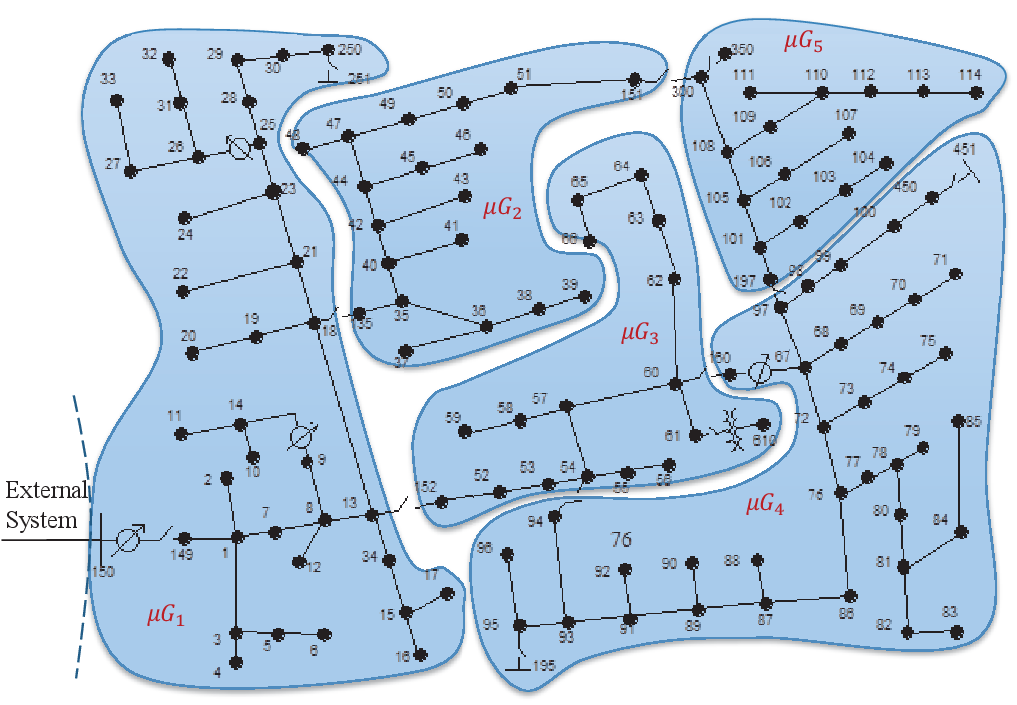}
		\caption{IEEE 123-feeder test network partitioned into five microgrids.}
		\label{fig:network}
		\vspace{-0.5em}
	\end{minipage}
	\hfill
	\begin{minipage}{0.53\textwidth}
		\centering
		\includegraphics[scale=0.44,trim=1.1cm 0.7cm 0.6cm 0cm]{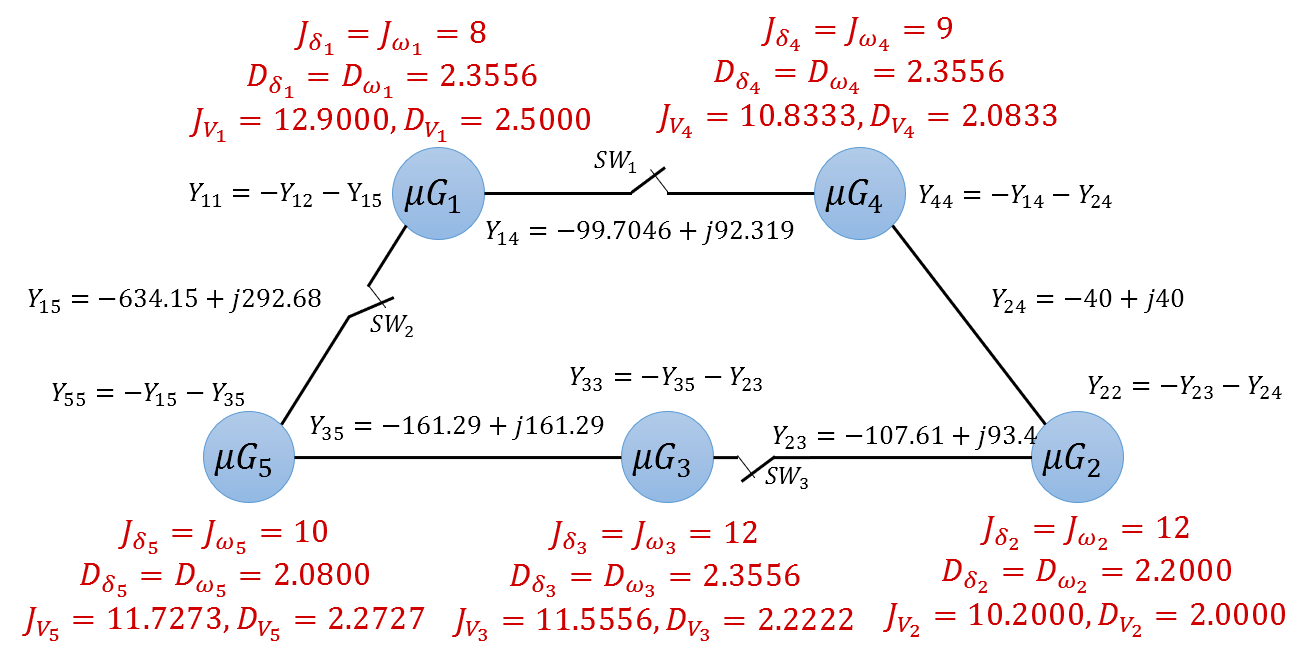}
		\caption{Network parameters (p.u.) for 123-feeder five-microgrid test system.}
		\label{fig:network_param}
		 \vspace{-0.5em}
	\end{minipage}
\end{figure*}

Using Theorem \ref{thm:design}, a distributed secondary controller can be designed for the microgrid interconnection with MAFD to guarantee stability even when D-PMU angle measurements are lost. { We note that Theorem \ref{thm:design} guarantees stability for arbitrary loss of D-PMU angle measurements at any microgrid, since the control design does not require any a priori knowledge of the switching vector $\sigma(t)$}. Further, by imposing a sparsity constraint \eqref{pass_lmi3} on the structure of the matrices $V_j$ and $U_j$, $j \in \Sigma$, we ensure that the secondary control law for microgrid $i$ only uses measurements from its immediate neighbors $\mathcal{N}_i$. Therefore, the sparsity structure of the distributed secondary controller $K_j$ will be the same as that of the network Jacobian matrix $H$, that is, $K_j \in \mathcal{S}_H$. { We also note that the D-MAFD control design can accommodate constraints on the power sharing between microgrids, which can either be incorporated while solving the power flow problem \eqref{powerflow} to provide the power reference set points, or as additional constraints limiting the range and/or rate of change of the control input $u_i(t)=[\Delta P_{inj}^i(t) \quad \Delta Q_{inj}^i(t)]'$ in the control design in Theorem \ref{thm:design}.}

\vspace{-1em}
\subsection{Robustness to network topology changes}
In microgrid interconnections, topology changes may frequently occur due to islanding of some microgrids or line outages. In such scenarios, it is important to ensure that the stability of the microgrid interconnection with the new topology can be guaranteed without redesigning the existing controllers in the system, even if D-PMU angle measurement losses occur during islanding or reconnection of microgrids. We accomplish this objective by incorporating a robustness margin into the D-MAFD secondary control design as follows.

Let the perturbation in the system Jacobian due to the change in network topology be given by
\vspace{-0.5em}
\setcounter{equation}{15}
\begin{equation}\label{deltaH}
\Delta H=H-H_{new},
\vspace{-0.5em}
\end{equation}
where $H_{new}$ is the Jacobian matrix after the network topology change. Then the stability of the microgrid interconnection with the new system topology with respect to disturbances $w(t)$ even under D-PMU measurement losses is guaranteed by the following robustness result.   
\begin{theorem}
	\label{thm:robustness} Given a network topology change with a new Jacobian matrix $H_{new}$, and $\Delta H$ as defined in \eqref{deltaH}, if there exists symmetric positive definite matrix \(P>0\), negative definite matrix $Q_j<0$, and matrices \(U_j, V_j, S_j\) and $R_j$ of appropriate dimensions such that \eqref{pert_lmi} is satisfied with $\gamma=||B_j^{(1)}\Delta H||_2 I$ for every $j \in \Sigma$, then the distributed control law $u(t) \mapsto u(t) + \tilde{u}(t)$ where \(\tilde{u}(t)=K_{\sigma(t)}y(t)\) with 
	\vspace{-0.25em}
	\setcounter{equation}{17}
	\begin{equation} 
	\label{gain1}
	K_j=V_j^{-1}U_j, \quad \forall j\in \Sigma,
	\vspace{-0.25em}
	\end{equation}
	guarantees that the closed loop dynamics \eqref{nonlinear_switched} is stable in the $\mathcal{L}_2$ sense with respect to any disturbance $w(t)$ for the new network topology. Furthermore, the control law $u(t) \mapsto u(t) + \tilde{u}(t)$ also guarantees stability of the closed loop system \eqref{nonlinear_switched} for any new network topology with Jacobian matrix $\hat{H}_{new}$ such that $||B_j^{(1)}\Delta \hat H||_2 I <\gamma$, where $\Delta \hat{H}=H-\hat{H}_{new}$.
\end{theorem}

\textit{Selection of robustness margin:} Theorem \ref{thm:robustness} presents a distributed secondary control design such that the microgrid interconnection in the D-MAFD framework is not only robust to a particular topology change, but also robust to any topology change that results in a smaller perturbation in the system Jacobian than the one used for the control design. Therefore, for maximal robustness, the D-MAFD secondary controller should be designed for the topology change that leads to the largest perturbation in the network Jacobian, that is, by selecting the robustness margin $\gamma$ to be the maximum value of $||B_j^{(1)}\Delta \hat H||_2$ over all possible topology changes. However, in practice, such a choice of $\gamma$ will require the computation of Jacobian matrices with respect to a very large number of network topologies, and may also be conservative. Therefore, the secondary control design using Theorem \ref{thm:robustness} can be carried out to guarantee $(N-1)$-robustness, that is, robustness in the scenario where a single microgrid is islanded or an outage takes place on a single line. In this case, an $(N-1)$-contingency analysis considering islanding or outage scenarios can be performed to select the worst-case robustness margin $\gamma$ for the secondary control synthesis. With this design, robustness of the microgrid interconnection to topology changes can be guaranteed even if D-PMU measurement losses occur during islanding, outages or restoration operations. 

\vspace{-1em}
\section{Case Studies}
\setcounter{figure}{4}
\begin{figure}
	\vspace{-0.5em}
	\centering
	\includegraphics[scale=0.58,trim=0.8cm 0.5cm 0cm 0.2cm]{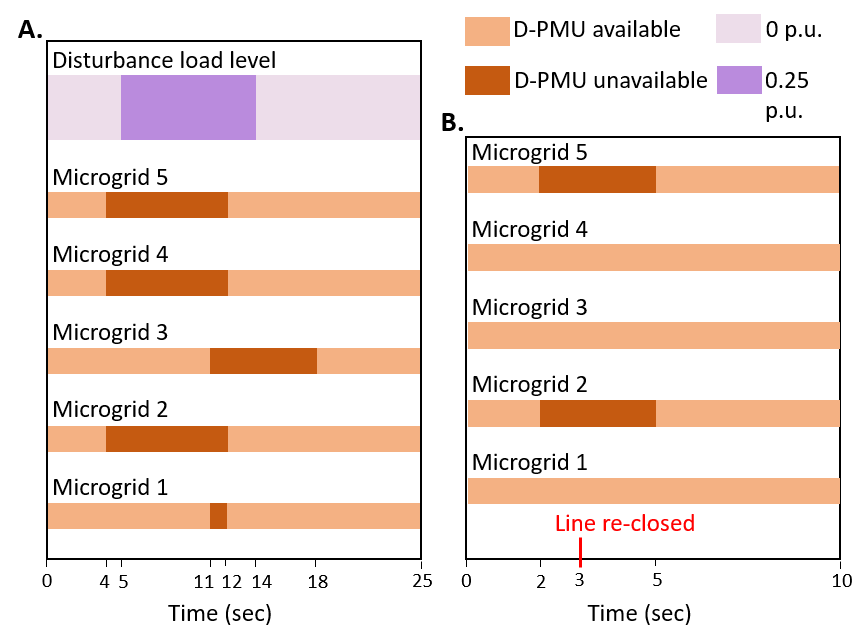}
	\caption{A. D-PMU measurement loss pattern and disturbance pattern for Scenario 1. B. D-PMU measurement loss pattern for Scenario 2.}
	\label{fig:case_switching}
	\vspace{0.6em}
	\includegraphics[scale=0.8,trim=0.1cm 0.5cm 0cm 0.2cm]{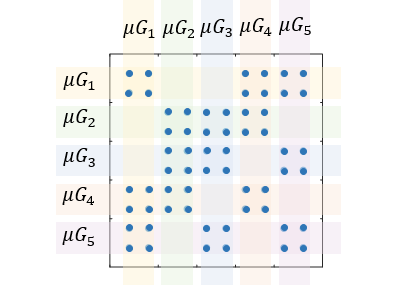}
	\caption{Sparsity structure of the D-MAFD secondary controller for five-microgrid test system.}
	\vspace{-1em}
	\label{fig:gain_saprsity}
\end{figure}
\setcounter{figure}{6}
\begin{figure*}[t]
	\centering
	\vspace{0.7em}
	\includegraphics[width=0.93\textwidth,trim=0.25cm 0.1cm 0cm 0.2cm]{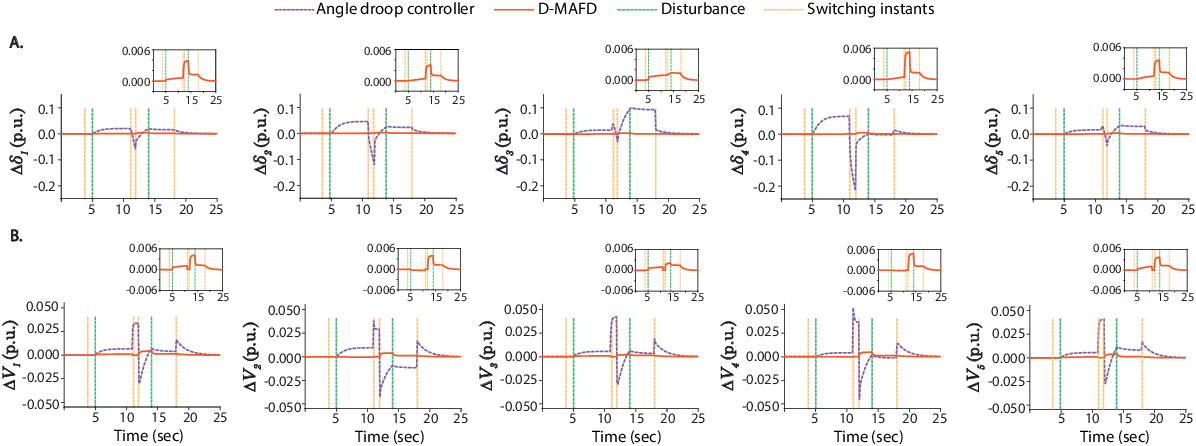}
		\vspace{-0.5em}
	\caption{A. Angle errors, and B. voltage errors of the D-MAFD design (\textbf{C1}) compared with a traditional angle droop controller (\textbf{C3}) for Scenario 1.}
	\label{fig:case}
	\vspace{1.5em}
	\includegraphics[width=0.93\textwidth,trim=0.1cm 0.1cm 0cm 0.2cm]{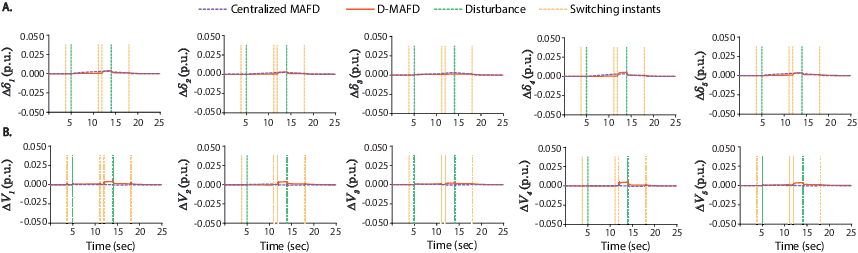}
	\vspace{-0.5em}
	\caption{A. Angle errors, and B. voltage errors of the D-MAFD design (\textbf{C1}) compared with a centralized secondary control design (\textbf{C2}) for Scenario 1.}
	\label{fig:cen_vs_dcen}
	 \vspace{-1em}
\end{figure*}

We demonstrate the performance of the D-MAFD control framework by considering a test five-microgrid interconnection (Fig. \ref{fig:network_param}) constructed as described in \cite{zhang2015online} from the 123-feeder test system shown in Fig. \ref{fig:network}. 
For this network with MAFD primary control, we obtain a nonlinear switched system model of the form \eqref{nonlinear_switched}. { Since every microgrid can switch between the angle droop control mode and the frequency droop control mode  at any time instant, the total number of switching modes of the system of five microgrids is given by $2^5=32$.} We then linearize the system around the power flow operating points in Table \ref{powerflow}. We present two test scenarios to illustrate the performance of the D-MAFD framework - (i)   under D-PMU measurement losses and disturbances, and (ii) during system topology changes due to line restoration after an outage.
\begin{table}[!t]
	\centering
	\vspace{-0.7em}
	\caption{Power flow solution for 123-feeder 5-microgrid test system}
	\label{operating_point_five}
	{\scriptsize
		\begin{tabular}{l|lllllll}
			\hline
			Condition & & $P_{inj}^{ref}$ & $Q_{inj}^{ref}$ & $P_{load}^{ref}$ & $Q_{load}^{ref}$ & $V^{ref}$ & $\delta^{ref}$    \\
			& & (p.u.) & (p.u.) & (p.u.) & (p.u.) & (p.u.) & (deg.)\\
			\hline
			& $\mu G_1$ & 0.79   &   1.35   &   0.92   &   0.47   & 1.000     & 0.000 \\
			(i) $SW_1$,& $\mu G_2$ & 0.80   &   0.10   &   0.23   &  0.11    & 1.003  &  0.233 \\
			$SW_2$, & $\mu G_3$ & 0.20   &  0.10    &  0.45    &  0.20     & 1.000 & 0.110 \\
			$SW_3$ & $\mu G_4$ & 0.80   &   0.20   &  0.27    &  0.12    & 1.003  &  0.158  \\
			closed & $\mu G_5$ & 0.20   &   0.10   &   0.92   &   0.95    & 0.999 &  0.052 \\\hline
			& $\mu G_1$ &  0.80  &  1.36     &  0.92    & 0.47     & 1.000     & 0.000 \\
			& $\mu G_2$ & 0.80  &  0.10   & 0.23     & 0.11     & 1.008  & 0.493   \\
			(ii) $SW_1$ & $\mu G_3$ &  0.20 & 0.10    &   0.45  & 0.20     & 1.002 & 0.227 \\
			open & $\mu G_4$ &  0.80  & 0.20   &    0.27  & 0.12    &  1.016 & 0.808  \\
			& $\mu G_5$ & 0.20 &  0.10    & 0.92    & 0.95    & 1.000 & 0.071  \\\hline
			& $\mu G_1$ & 0.82  & 1.38  &  0.92    &  0.47   & 1.000     & 0.000 \\
			& $\mu G_2$ &  0.80  & 0.10     & 0.23   & 0.11    &  0.978 & 0.727  \\
			(iii) $SW_2$ & $\mu G_3$ & 0.20  & 0.10      &    0.45  & 0.20      & 0.968 & 0.763 \\
			open & $\mu G_4$ & 0.80  &  0.20   &    0.27 & 0.12   & 0.996 & 0.312   \\
			& $\mu G_5$ &  0.20  & 0.10   &  0.92 &    0.95   & 0.963 & 0.788 \\\hline
			& $\mu G_1$ &  0.80 & 1.36   & 0.92  & 0.47   &    1.000  & 0.000 \\
			& $\mu G_2$ & 0.80  &  0.10   & 0.23 & 0.11     & 1.013  & 0.699  \\
			(iv) $SW_3$ & $\mu G_3$ &  0.20  & 0.10 &  0.45  & 0.20     & 0.997 & 0.012\\
			open & $\mu G_4$ & 0.80  & 0.20     &  0.27  & 0.12    & 1.006  & 0.292   \\
			& $\mu G_5$ & 0.20 &  0.10    &    0.92 & 0.95   & 0.998 & 0.038  \\\hline
	\end{tabular}}			\vspace{-1.2em}
\end{table}
\setcounter{figure}{7}
\vspace{-1.2em}
\subsection{Scenario 1: D-PMU measurement loss and load change}
Here, we assess the performance of the D-MAFD framework under D-PMU measurement losses. We use the linearized system model around the power flow solution for Condition (i) in Table \ref{powerflow} to design three controllers for comparison:
\begin{itemize}
	\item \textbf{C1:} a distributed output-feedback secondary controller (D-MAFD controller) by solving \eqref{control_lmi}-\eqref{gain} for every $j \in \Sigma$,
	\item \textbf{C2:} a centralized output-feedback secondary controller by solving \eqref{pass_lmi1}, \eqref{pass_lmi2}, \eqref{gain} for every $j \in \Sigma$, and
	\item \textbf{C3:} a centralized secondary controller by solving \eqref{pass_lmi1}, \eqref{pass_lmi2}, \eqref{gain} for the microgrid interconnection where all microgrids continue to use angle droop control with the last available measurement even when D-PMU measurements are lost, that is, with the dynamics corresponding to the mode $j=[1 \, 1 \ldots \, 1]$ in \eqref{nonlinear_switched}.
\end{itemize}
The control design LMIs are solved using YALMIP \cite{lofberg2004yalmip}. We consider a test pattern of D-PMU measurement losses and a disturbance acting on all microgrids as shown in Fig. \ref{fig:case_switching}-A. For this test pattern, we simulate the nonlinear system dynamics with each of the designed controllers and make the following observations about their performance:

\begin{itemize}
	\item We note that the sparsity structure of the D-MAFD secondary controller (Fig. \ref{fig:gain_saprsity}) is the same as that of the network Jacobian, indicating that each microgrid only uses output measurements from its immediate neighbors. 
	\item From the angle and voltage profiles in Fig. \ref{fig:case}, we observe that the D-MAFD control design is successful in stabilizing the microgrid interconnection under the test D-PMU measurement loss scenario in the presence of disturbances. On the other hand, when all microgrids continue to use angle droop control with the last available measurement during D-PMU measurement loss, the system suffers from poor transient performance, and the angle and voltage droop errors continue to increase from $t=12s$ until the disturbance is withdrawn at $t=14s$, indicating that angle droop control alone is unable to stabilize the system in this scenario. 
	\item A comparison with the performance of the centralized secondary control design for this scenario indicates that the performance of the D-MAFD design is comparable (Fig. \ref{fig:cen_vs_dcen}), despite the secondary controller in the D-MAFD framework using limited information from other microgrids in the network. This is an advantage, since distributed control designs are typically significantly outperformed by centralized designs. 
	\item { 
		The time constant of the D-MAFD controller is slightly higher, that is, the D-MAFD controller has slightly slower dynamics than the traditional angle droop controller. This is natural, since  indirect control of the angle via frequency measurements is carried out in the D-MAFD framework during angle measurement loss. Therefore, the D-MAFD controller, while being much  faster than a traditional frequency droop controller, is still slightly slower than the angle droop controller.  However, we note that the angle droop controller cannot stabilize the network under loss of D-PMU angle measurements, while the D-MAFD controller can stabilize the network under arbitrary measurement losses, as seen in Fig. \ref{fig:case}.}
	\item { As mentioned in Remark \ref{rem:continuity}, the MAFD framework allows for seamless switching between the angle and frequency droop control modes when angle measurements are lost or restored. This is confirmed by the continuity of the state and the absence of any switching transients in our case study in Fig. \ref{fig:case}.}
\end{itemize}

\vspace{-1.2em}
\subsection{Scenario 2: Line reclosing after outage}
In order to demonstrate the robustness of the D-MAFD control design to changes in topology, we consider three fault scenarios that result in a line outage due to the opening of either $SW_1$, $SW_2$ or $SW_3$ in Fig. \ref{fig:network_param}. The power flow solutions for each case are provided in Table \ref{powerflow}. From a contingency analysis of the system, we determine that the opening of $SW_2$ (outage on the tie line between microgrids $\mu G_1$ and $\mu G_5$) results in the worst case $\gamma$ in Theorem \ref{thm:robustness}. We design the D-MAFD controller corresponding to this outage by solving \eqref{pert_lmi}. 

We study the performance of this D-MAFD control design when the faulted line is restored by reclosing $SW_2$ at $t=5s$, under the D-PMU measurement loss scenario shown in Fig. \ref{fig:case_switching}-B. Prior to the reclosing operation, the system initial condition corresponds to Condition (iii) in Table \ref{powerflow}. After the reclosing operation, it is desired that the system returns to the original operating point corresponding to Condition (i) in Table \ref{powerflow}. From the angle and voltage error profiles for this scenario (Fig. \ref{fig:fault}-B), we observe that the D-MAFD control design maintains system stability even when a D-PMU measurement loss occurs during the reclosing operation. 

We also evaluate the robustness of the designed controller by evaluating its performance for two other scenarios where $SW_1$ (Fig. \ref{fig:fault}-A) and $SW_3$ (Fig. \ref{fig:fault}-C) are reclosed after outages resulting from faults. We observe that the D-MAFD secondary controller designed for the worst-case line outage configuration ($SW_2$ open) is also successful in maintaining system stability in these two scenarios. This indicates that the D-MAFD framework is robust to network topology changes and does not require redesigning the secondary controller to maintain stability under different interconnection topologies. 

\begin{figure}
	\setcounter{figure}{8}
	\centering
	\vspace{1em}
	\includegraphics[scale=0.98,trim=0cm 0.3cm 0cm 0.2cm]{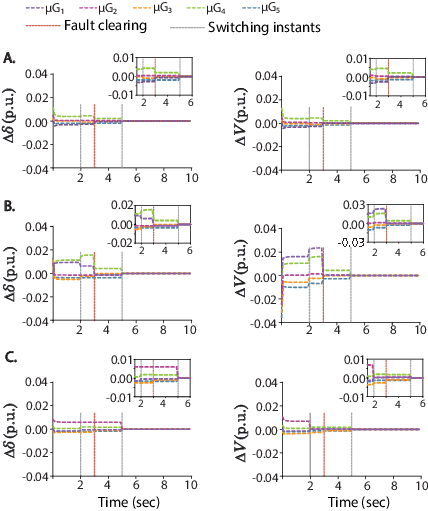}
	\caption{Angle and voltage errors of the D-MAFD design for Scenario 2, reclosing A. $SW_1$, B. $SW_2$ and C. $SW_3$.}
	\vspace{-1em}
	\label{fig:fault}
\end{figure}

\vspace{-0.6em}
\section{Conclusion}
We presented a mixed voltage angle and frequency droop control framework with a distributed secondary controller (D-MAFD framework) for distribution-level microgrid interconnections where loss of D-PMU angle measurements may result in degradation of stability and performance. The proposed D-MAFD framework provides a stable and robust solution to enhance the reliability of D-PMU based control designs for microgrid interconnections. In addition to D-PMU measurement loss scenarios, the D-MAFD framework can also be applied to legacy systems in which some microgrids operate with traditional frequency droop control and others employ newer angle droop control schemes.

\vspace{-0.5em}
\section*{Appendix}

{ \noindent Here, we present the proofs of all the results in Section \ref{sec:dmafd}. }

\vspace{0.5em}
\noindent \textbf{Proof of Proposition \ref{prop:linear_to_nonlinear}:} If  system \eqref{linear switched system} is $QSR$-SSID, then \eqref{QSR-SSID} is true for all $j\in\Sigma$. Substituting \eqref{linear switched system} in \eqref{QSR-SSID}, 
{{\begin{align}\label{Prop1_linear}
		\begin{bmatrix}
		x(t) \\ w(t)
		\end{bmatrix}'\Gamma^{(j)}\begin{bmatrix}
		x(t) \\ w(t)
		\end{bmatrix} > \phi_j(w(t)) + \psi_j(x(t)),
		\end{align}}}where {\small{$\Gamma^{(j)}_{11}{=}C_j'Q_jC_j{-}P(A_j+B_j^{(1)}H){-}(A_j+B_j^{(1)}H)'P$, $\Gamma^{(j)}_{12}=C_j'Q_jD_j+C_j'S_j-PB_j^{(2)}, \, \Gamma^{(j)}_{21}=\Gamma_{12}^{(j)^{'}}$}}, {\normalsize and} {\small$\Gamma^{(j)}_{22}=D_j'Q_jD_j+D_j'S_j+S_j'D_j+R_j$}. Now consider 
{\small
	\begin{multline}\label{Prop1_nonlinear}
	\Lambda_j = \begin{bmatrix}
	y(t) \\
	w(t)
	\end{bmatrix}'\begin{bmatrix}
	Q_j & S_j\\
	S_j' & R_j
	\end{bmatrix}\begin{bmatrix}
	y(t) \\
	w(t)
	\end{bmatrix} - \dot{V}(x(t))  \\ - \phi_j(w(t)) - \psi_j(x(t)), \, j \in \Sigma, 
	\end{multline}}for the nonlinear switched system \eqref{nonlinear_switched}. Since the linearization \eqref{linear switched system} is obtained by computing a first order Taylor approximation of every mode of \eqref{nonlinear_switched}, 
we can substitute for $\dot{x}(t)$ and $y(t)$ from \eqref{nonlinear_switched} in \eqref{Prop1_nonlinear} and write the Taylor series expansions of $f_j$, $g_j$ and $h$ around $x=0$ and $w=0$. Using \eqref{Prop1_linear}, we can then show that $\Lambda_j$ is upper bounded by a function of the higher order terms in the Taylor series expansion. Then, along the lines of \cite[Theorem 3.1]{wang2018passivity}, these higher order terms can be upper bounded  to show that $\Lambda_j>0$ in a neighborhood of $x=0, \, w=0$ for all $j \in \Sigma$, completing the proof. \hspace*{\fill}$\blacksquare$

\vspace{1em}

\noindent \textbf{Proof of Theorem \ref{thm:design}:}
The dynamics of closed loop system \eqref{linear switched system} with output feedback controller $u(t) \mapsto u(t) + \tilde{u}(t)$, \(\tilde{u}(t)=K_{\sigma(t)}y(t)\) are given by 
\vspace{-0.5em}
\begin{subequations}\label{linear closed loop}
	{\small{\begin{align*}
			\dot{{x}}(t) = \hat{A}_{\sigma(t)}{x}(t) + \hat{B}_{\sigma(t)}^{(2)}{w}(t), \;
			{y}(t)  = \hat{C}_{\sigma(t)}{x}(t) + \hat{D}_{\sigma(t)}{w}(t), 
			\end{align*}}}
\end{subequations}where {{$\hat{A}_{\sigma(t)} = A_{\sigma(t)}+B^{(1)}_{\sigma(t)}H+B^{(1)}_{\sigma(t)}K_{\sigma(t)}C_{\sigma(t)}$, $\hat{B}_{\sigma(t)}^{(2)}=B_{\sigma(t)}^{(2)}+B^{(1)}_{\sigma(t)}K_{\sigma(t)}D_{\sigma(t)}$, $\hat{C}_{\sigma(t)}=C_{\sigma(t)}$}} and {{$\hat{D}_{\sigma(t)}=D_{\sigma(t)}$}}.  Since {{$P>0$}}, it is full rank. If matrices {{$B_j^{(1)}$, $j \in \Sigma$}} are full column rank, then $V_j$ satisfying \eqref{pass_lmi2} are invertible. Substituting equations \eqref{gain} and \eqref{pass_lmi2} in \eqref{pass_lmi1} gives 
\setcounter{equation}{20}
{\small\begin{equation}
	\begin{bmatrix}
	-P\hat{A}_j-\hat{A}_j'P & \hat{C}_j'S_j-P\hat{B}_j^{(2)} & -\hat{C}_j'Q_{j-}^{1/2} \\
	S_j'\hat{C}_j-\hat{B}_j^{(2)'}P & \hat{D}_j'S_j+S_j'\hat{D}_j+R_j & -\hat{D}_j'Q_{j-}^{1/2}\\
	-Q_{j-}^{1/2}\hat{C}_j & -Q_{j-}^{1/2}\hat{D}_j & I
	\end{bmatrix}>0,
	\end{equation}}$\forall j \in \Sigma$, where {{$Q_{j-}^{1/2} Q_{j-}^{1/2} = -Q_j$}}. By taking the Schur's complement, it is easy to conclude that the closed loop system \eqref{linear switched system} with the controller in Theorem \ref{thm:design} is $QSR$-SSID. The result can now be obtained using Proposition \ref{prop:linear_to_nonlinear} and Remark \ref{remark: QSR-L2}. \hspace*{\fill}$\blacksquare$

\vspace{1em}

\noindent \textbf{Proof of Theorem \ref{thm:robustness}:}
Consider the closed loop system \eqref{nonlinear_switched} with a new Jacobian matrix $H_{new}$ and the control law $u(t) \mapsto u(t) + \tilde{u}(t)$, where \(\tilde{u}(t)=K_{\sigma(t)}y(t)\), $\forall j\in\Sigma$ is obtained from \eqref{pert_lmi}-\eqref{gain1}. Now consider the matrix $M_j$ in \eqref{pass_lmi1}  with its first term updated to {{$[M_j]_{11} = -P(A_j+B_j^{(1)}H_{new})-(A_j+B_j^{(1)}H_{new})'P 
		-B_j^{(1)}U_jC_j-C_j'U_j'B_j^{(1)'}$}}.
Then,
\footnotesize \begin{align}
M_j - \hat{M}_j = \begin{bmatrix}[1.5]
-P(B_j^{(1)}\Delta H)-(B_j^{(1)}\Delta H)'P + 2\gamma P & 0 & 0 \\
0 & 0 & 0 \\
0 & 0 & 0
\end{bmatrix},
\end{align}
\normalsize
where {{$\gamma =||B_j^{(1)}\Delta H||_2 I $}}. Clearly, since {{$\gamma \geq B_j^{(1)}\Delta H$, $ M_j - \hat{M}_j \geq 0$}}. If \eqref{pert_lmi} holds, {{$ M_j \geq \hat{M}_j >0$ $\implies$ $M_j>0$}}. Thus, using Theorem \ref{thm:design}, if \eqref{pert_lmi}, \eqref{gain1} holds, the closed loop system \eqref{nonlinear_switched} is locally $\mathcal{L}_2$ stable with the new Jacobian matrix {{$H_{new} = H + \Delta H$}}. It is then fairly straightforward to show that this control law renders the closed loop system $\mathcal{L}_2$-stable for any new network topology with Jacobian matrix {{$\hat{H}_{new}$}} such that {{$||B_j^{(1)}\Delta \hat H||_2 I <\gamma$}}, where {{$\Delta \hat{H}=H-\hat{H}_{new}$}}. \hspace*{\fill}$\blacksquare$
\section*{Acknowledgment}
The first author thanks Shailaja Seetharaman for the design of the illustrations  in Figs. \ref{fig:microgrids_schematic} and \ref{fig:dmafd_schematic}.

\bibliographystyle{IEEEtran}
\bibliography{references}

\begin{IEEEbiography}[{\includegraphics[width=1in,height=1.25in,clip,keepaspectratio]{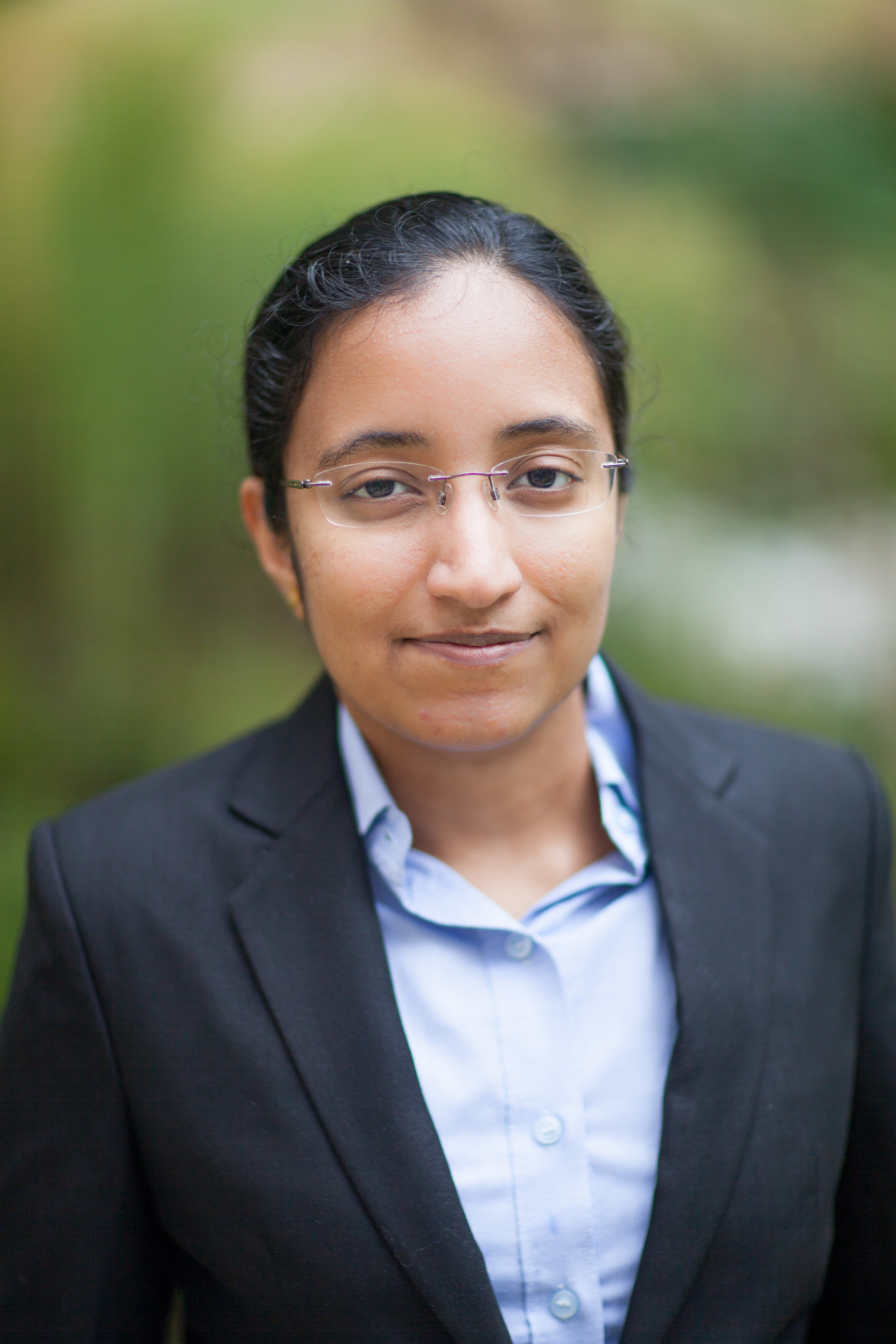}}]{S Sivaranjani} is a postdoctoral researcher in the Department of Electrical and Computer Engineering at Texas A\&M University. She
	received her Ph.D., M.S., and B.E. degrees, all in Electrical Engineering, from the University of Notre Dame, the Indian Institute of Science, and the PES Institute of Technology, respectively. Her research interests are in data-driven and distributed control for large-scale infrastructure networks, with emphasis on energy systems and transportation networks. She is a recipient of the Schlumberger Foundation Faculty for the Future fellowship (2015-18), the Zonta International Amelia Earhart fellowship (2015-16) and the Notre Dame Ethical Leaders in STEM fellowship (2016-17).
\end{IEEEbiography}
\vspace{-2em}
\begin{IEEEbiography}[{\includegraphics[width=1in,height=1.25in,clip,keepaspectratio]{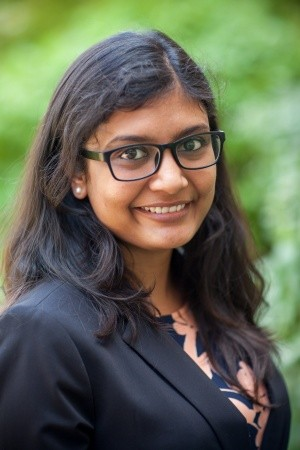}}]{Etika Agarwal}
	received her Ph.D. and M.S. degrees, both in Electrical Engineering, from the University of Notre Dame in 2019 and 2016 respectively, and her B. Tech in Avionics from the Indian Institute of Space Science and Technology in 2012. Before joining the graduate school, she worked with the Indian Space Research Organization from 2012-2014. Her research interests are in computationally efficient and scalable control of large-scale cyber-physical systems. 
\end{IEEEbiography}

\vspace{-2em}
\begin{IEEEbiography}[{\includegraphics[width=1in,height=1.25in,clip,keepaspectratio]{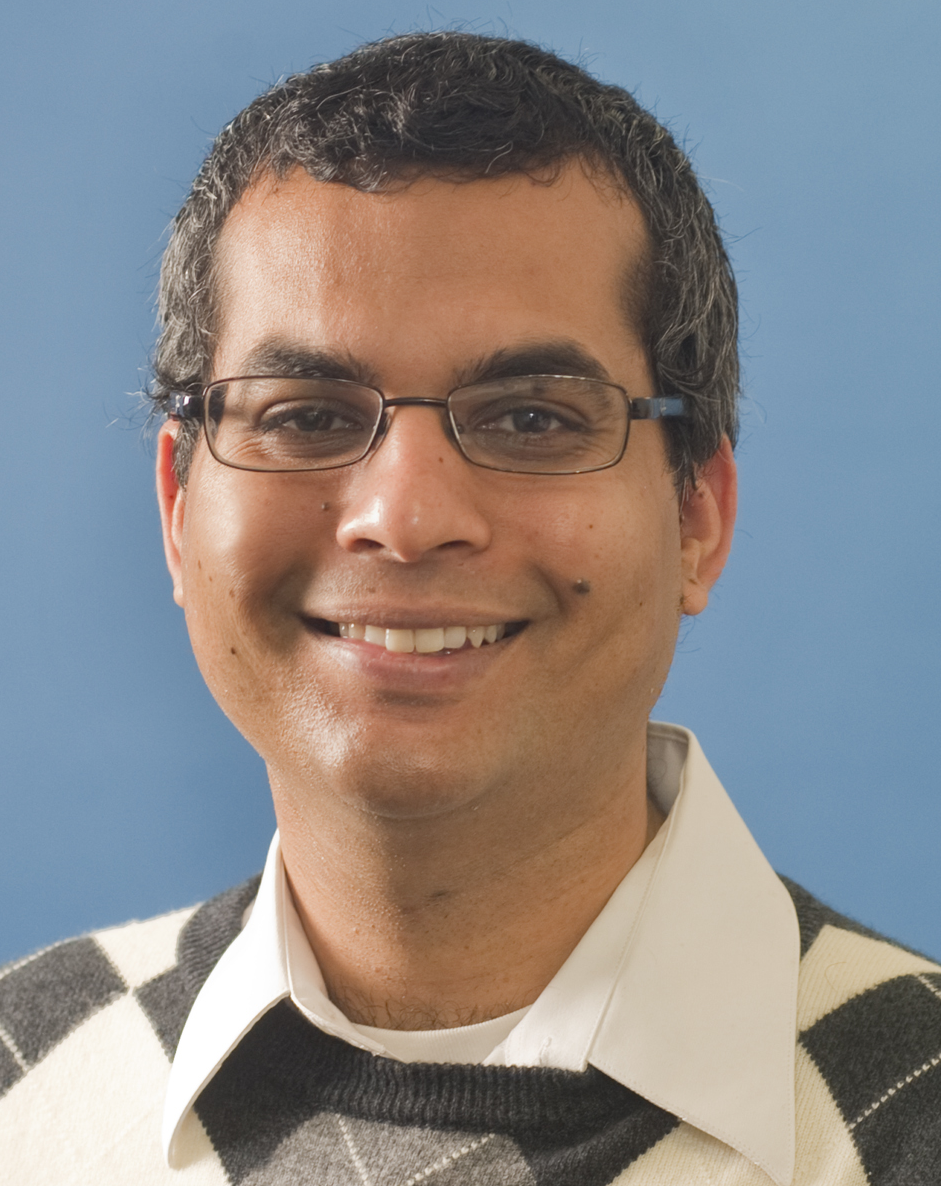}}]{Vijay Gupta} Vijay Gupta is a Professor in the Department of Electrical Engineering at the University of Notre Dame, having joined the faculty in January 2008. He received his B. Tech degree at Indian Institute of Technology, Delhi, and his M.S. and Ph.D. at California Institute of Technology, all in Electrical Engineering. He received the 2018 Antonio J Rubert Award from the IEEE Control Systems Society, the 2013 Donald P. Eckman Award from the American Automatic Control Council and a 2009 National Science Foundation (NSF) CAREER Award. His research and teaching interests are broadly in the interface of communication, control, distributed computation, and human decision making.
\end{IEEEbiography}

\begin{IEEEbiography}[{\includegraphics[width=1in,height=1.25in,clip,keepaspectratio]{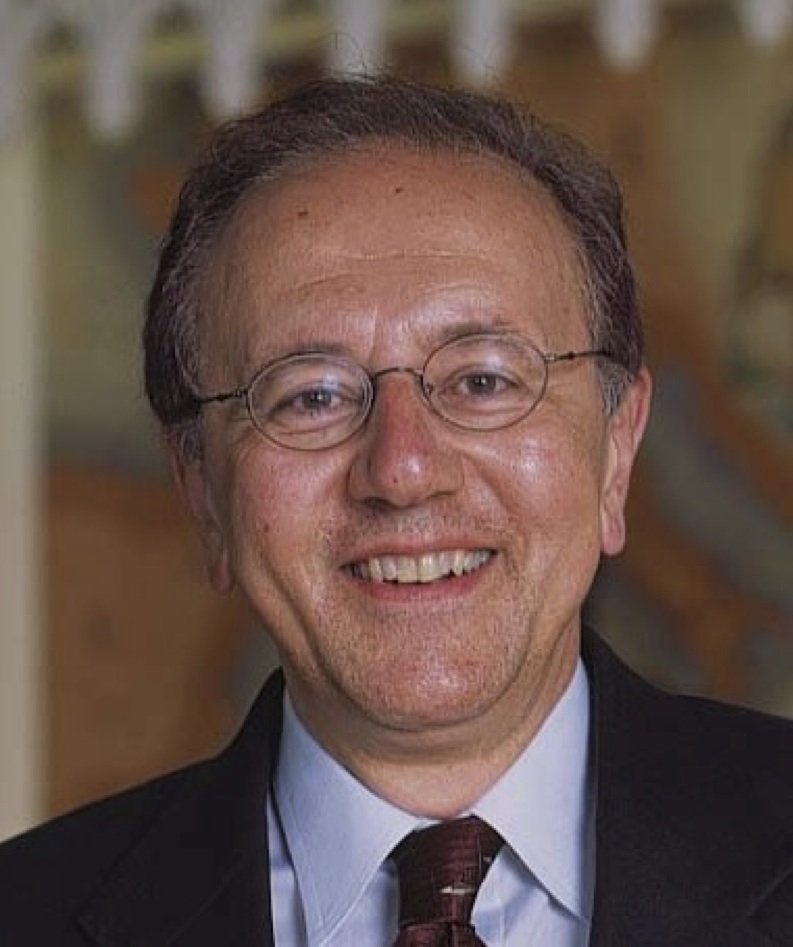}}]{Panos Antsaklis} is the H.C. \& E.A. Brosey Professor of Electrical Engineering at the University of Notre Dame. He is graduate of the National Technical University of Athens, Greece, and holds MS and PhD degrees from Brown University. His research addresses problems of control and automation and examines ways to design control systems that will exhibit high degree of autonomy. His current research focuses on Cyber-Physical Systems and the interdisciplinary research area of control, computing and communication networks, and on hybrid and discrete event dynamical systems. He is IEEE, IFAC and AAAS Fellow, President of the Mediterranean Control Association, the 2006 recipient of the Engineering Alumni Medal of Brown University and holds an Honorary Doctorate from the University of Lorraine in France. He served as the President of the IEEE Control Systems Society in 1997 and was the Editor-in-Chief of the IEEE Transactions on Automatic Control for 8 years, 2010-2017.
\end{IEEEbiography}

\vspace{-42em}
\begin{IEEEbiography}[{\includegraphics[width=1in,height=1.25in,clip,keepaspectratio]{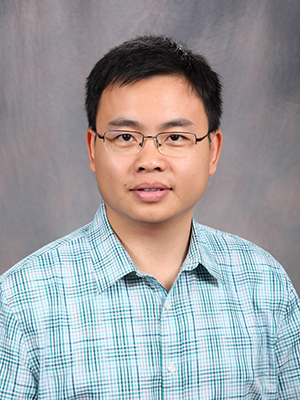}}]{Le Xie}(S'05-M'10-SM'16) received the B.E. degree in electrical engineering from Tsinghua University, Beijing, China, in 2004, the M.S. degree in engineering sciences from Harvard University, Cambridge, MA, USA, in 2005, and the Ph.D. degree from the Department of Electrical and Computer Engineering, Carnegie Mellon University, Pittsburgh, PA, USA, in 2009. He is currently a Professor with the Department of Electrical and Computer Engineering, Texas A\&M University, College Station, TX, USA. His research interests include modeling and control of large-scale complex systems, smart grids application with renewable energy resources, and electricity markets.
\end{IEEEbiography}

\end{document}